\documentclass[article]{elsarticle}

\usepackage{geometry}
\usepackage{balance}
\usepackage{multicol}
\usepackage{amssymb}
\usepackage{amsmath}
\usepackage{amsfonts}
\usepackage{booktabs}
\usepackage{graphicx}
\usepackage{lineno,hyperref}
\modulolinenumbers[100]
\tiny
\journal{\tiny Nuclear Instruments and Methods in Physics Research Section A:
Accelerators, Spectrometers, Detectors and Associated Equipment}

\begin{document}\small


\title{Recovery of Saturated $\gamma$ Signal Waveforms by Artificial Neural Networks}

%

\author[mymainaddress]{Yu Liu}

\author[mysecondaryaddress]{Jing-Jun Zhu}

\author[mythirdaddress]{Neil Roberts}

\author[mymainaddress]{Ke-Ming Chen}

\author[mymainaddress]{Yu-Lu Yan}

\author[mysecondaryaddress]{Shuang-Rong Mo}

\author[mymainaddress]{Peng Gu}

\author[mymainaddress]{Hao-Yang Xing\corref{mycorrespondingauthor}}

\cortext[mycorrespondingauthor]{Corresponding author}
\ead{xhy@scu.edu.cn}

\address[mymainaddress]{School of Physical Science and Technology, Sichuan University, Chengdu, 610064, China}
\address[mysecondaryaddress]{Key Laboratory of Radiation Physics and Technology of Ministry of Education,
Institute of Nuclear Science and Technology, Sichuan University, Chengdu, 610065, China}
\address[mythirdaddress]{Edinburgh Imaging, School of Clinical Sciences, University of Edinburgh, United Kingdom}

\begin{frontmatter}

\begin{abstract}

 Particle may sometimes have energy outside the range of radiation detection hardware so that the signal is saturated and useful information is lost. We have therefore investigated the possibility of using an Artificial Neural Network (ANN) to restore the saturated waveforms of $\gamma$ signals. Several ANNs were tested, namely the Back Propagation (BP), Simple Recurrent (Elman), Radical Basis Function (RBF) and Generalized Radial Basis Function (GRBF) neural networks (NNs) and compared with the fitting method based on the Marrone model. The GBRFNN was found to perform best.

\end{abstract}

\tiny

\end{frontmatter}

\linenumbers

\section{Introduction}

In particle physics experiments useful information is frequently lost because signals have energy outside the range of radiation detection hardware. For example, in an experiment which aims to detect dark matter, it is possible that many of the background signals will have an energy greater that can be detected. Restoring this signal and thus better characterizing the background radiation will provide favorable conditions for the detection of dark matter. To our knowledge, there has been relatively little research concerning the recovery of saturated signals and this is perhaps surprising since many models have been reported which provide a good fit for the waveforms of $\gamma$ and neutron radiation. Artificial Neural Networks (ANNs) have previously been used to discriminate between neutrons and $\gamma$-rays in a liquid scintillation detector \cite{Ref_1}, to reconstruct the neutron spectrum \cite{Ref_2} and to optimize the concrete mixture that will provide a thermal neutron shield \cite{Ref_3}. Here we investigate whether ANNs can also be used to successfully recover saturated signals in particle detectors. Initial analysis using simple waveform enlarging and model fitting of the gamma waveforms generated from a liquid scintillation detector was not able to recover the saturated signals. However, an excellent result was obtained by using the GRBFNN.

\section{Background}

\subsection{Waveform normalization}

For most particle signal detection systems the output signal waveform and the energy of the incident particles are assumed to be approximately linear and the shape of the signal uniform after normalization. However, in liquid scintillation detectors, as the energy of the particles increases the normalized output waveform will broaden. This non-linear behavior makes it difficult to recover the saturated signal by simply scaling.

\begin{figure}
 \centering
 \includegraphics[scale=0.5]{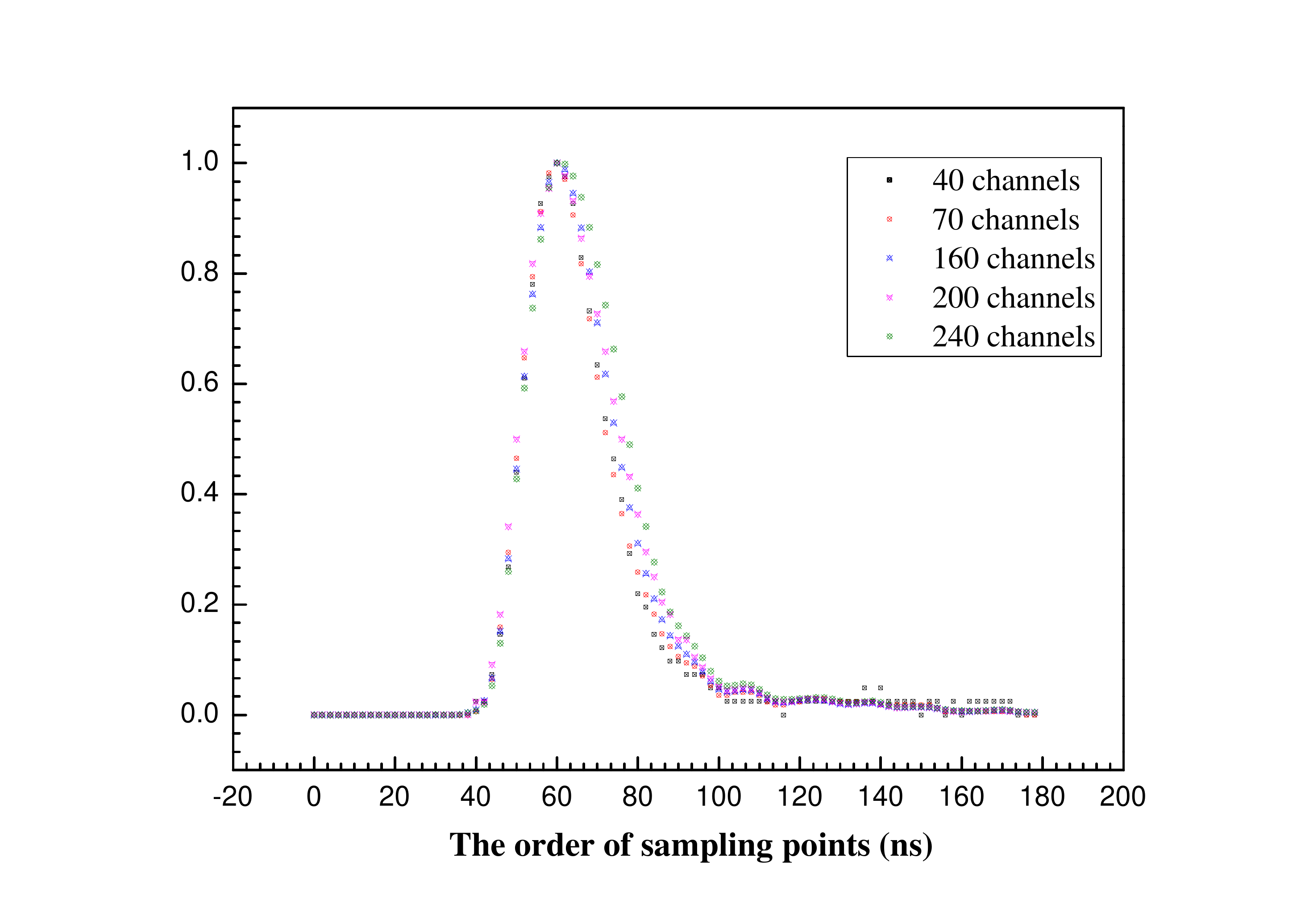}
 \caption{Comparison of normalized ¦Ã signal waveforms with different energy.}\label{normalize_p}
\end{figure}

The differences between gamma signals with different energy are illustrated in Figs. \ref{normalize_p}. In particular, the black, red, blue, purple and green dots are the normalized sampling points of the $\gamma$ signals with energy of 40, 70, 160, 200 and 240 channels and it can be seen that the falling edge goes linear in the lower energy range but is no longer linear in the higher energy range.

\subsection{Model fitting}

Although not completely consistent after normalization, the waveforms in Figs. \ref{normalize_p} are regular. If it is possible to model the waveform and extract relevant features, then it may be possible to use the behavior at different energies to infer the fitting parameters for the higher energy waveforms and thus simulate the waveform of the higher energy $\gamma$ signal and recover the waveform of the saturated gamma signal.

The model proposed by Marrone \emph{et al.}\cite{Ref_4} shown in Eq. \ref{Equation_1} has been shown to provide a good fit to signals from liquid scintillation detectors.

\begin{equation}\label{Equation_1}
L=A(exp^{-\theta(t-t_{0})}-exp^{-\lambda_{s}(t-t_{0})})+B(exp^{-\theta(t-t_{0})}-exp^{-\lambda_{l}(t-t_{0})})
\end{equation}

However, there are challenges in its verification such that no matter what initial parameters are selected, there is always a big difference between the fitted and  actual $\gamma$ signal waveforms and large uncertainty in the fitted parameters. Thus it is almost impossible to use a simple mathematical model to describe the waveform of the $\gamma$ signal generated by all liquid scintillation detectors with different hardware or liquid scintillation.

\begin{figure}
 \centering
 \includegraphics[scale=0.5]{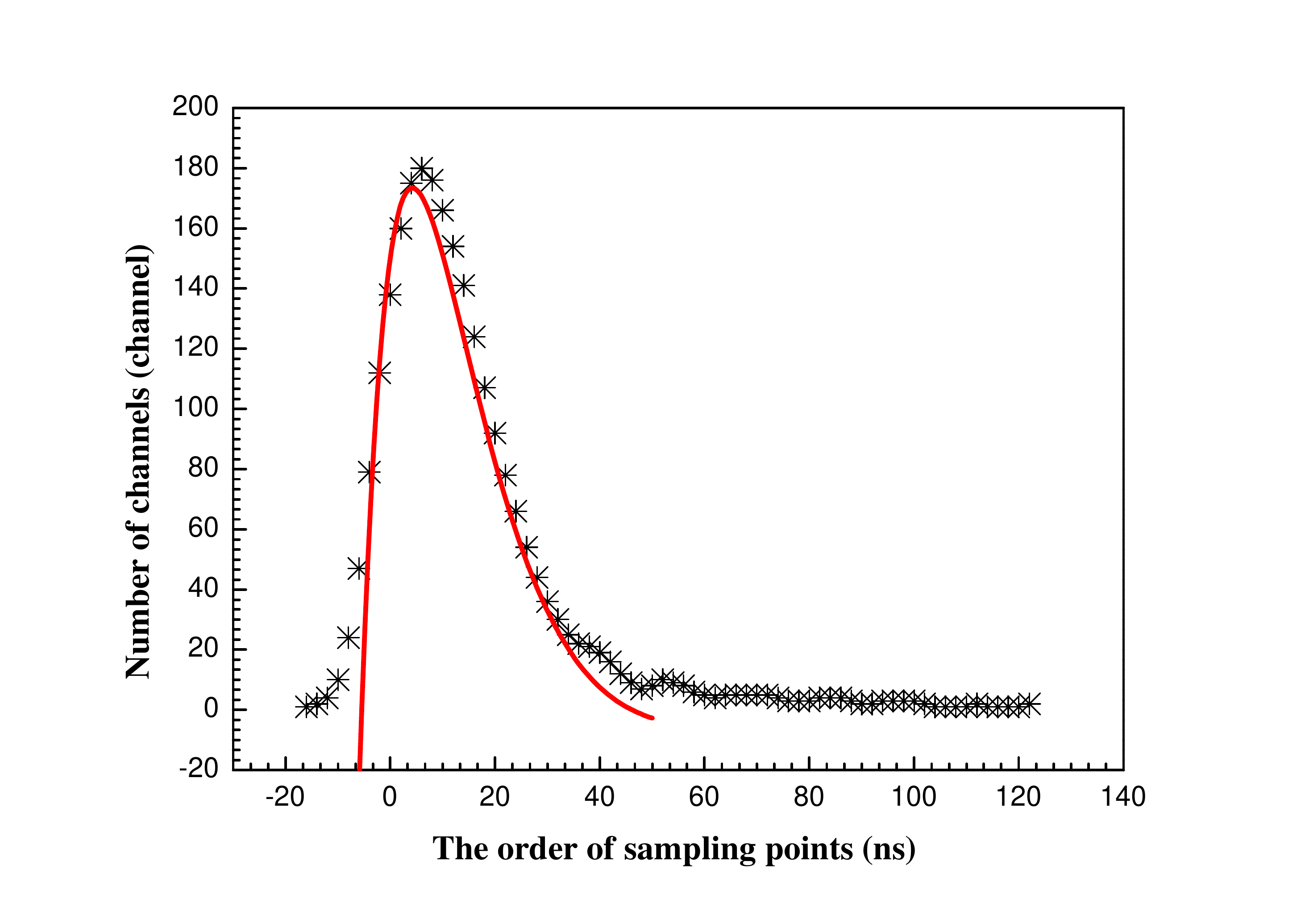}\\
 \caption{Model fitting by Marrone model.}\label{fit_p}
\end{figure}

The locally amplified $\gamma$ signal waveform and Marrone model fitting curve are shown in Figs. \ref{fit_p}. The black points are representative of the sampling of the $\gamma$ signal waveform and the red line is the fitted curve. The rising and falling edges of the $\gamma$ signal are not well fitted.

\section{Methodology}

ANNs are suitable for approximating complex relationships between input and output variables including non-linearity optimization \cite{Ref_5}. Therefore, saturated $\gamma$ signal waveforms with continuous energy distribution can be used as input and the corresponding complete waveforms as output to train the ANN which can find the relationship and recover the higher energy saturated $\gamma$ signal waveforms automatically.

\subsection{Construction of training set and test set in artificial neural network}

\begin{figure}
 \centering
 \includegraphics[scale=0.5]{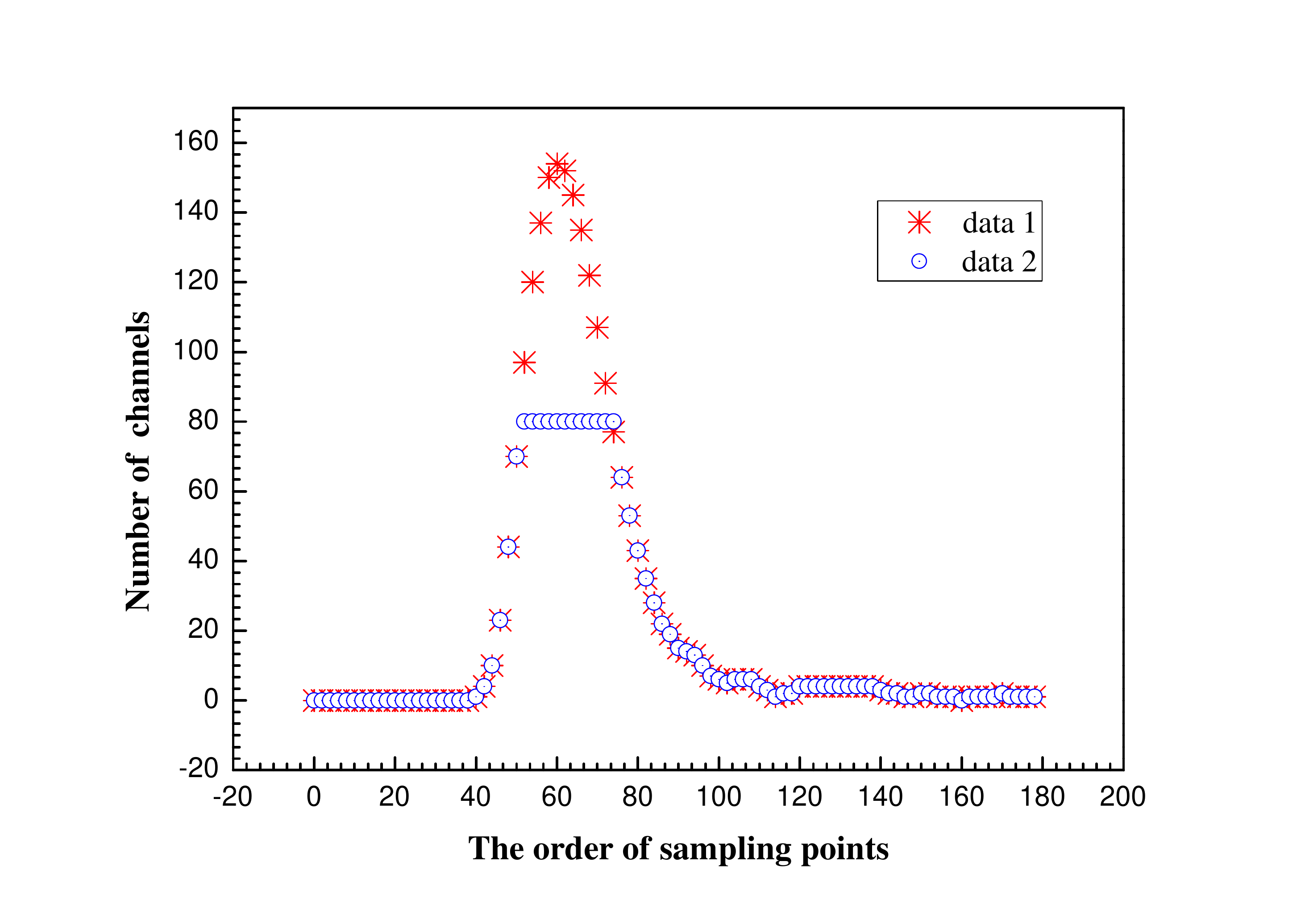}\\
 \caption{Comparison of normalized $\gamma$ signal waveforms with different energy.}\label{cutoff_p}
\end{figure}

The construction of the training and testing set is extremely critical. Since we cannot know the form of the true saturated waveform the approach adopted in this study has been to take away the signal in the peak region of standard waveforms and to replace it with a horizontal linear trace. Therefore, the complete waveforms recovered by the ANN¡¯s can be compared with the original waveforms to assess the error of the result and to determine performance. An example is given in Figs. \ref{cutoff_p} where the red symbols (data1) denote a complete gamma waveform and the blue symbols (data2) denote the corresponding simulated saturated waveform.

The original background data set was obtained using a EJ335 liquid scintillation detector in the laboratory of the CEDX experimental group of Sichuan University under Mount Jingping. A wide range of different types of particle signals with a variety of shapes are detected and n and ¦Á signals and other interference signals were removed prior to performing the present study. In addition, since the $\gamma$ signals in the low energy range cannot be properly distinguished from noise they were also eliminated. The remaining signals are used to define the energy distribution in the training set and which is lower than the energy distribution in the test set.

\subsection{Cutoff value (saturation value) setting of input data}

The proportion of saturated signals in the energy distribution is not known and so we a random value in the range 4\% to 99\% is taken as the saturation value in both the training and testing datasets at first. Then the saturation value of the true saturated waveform is actually determined and the randomness of the ratio between the energy and the saturation value is determined by the randomness of the energy were considered. Choosing a certain value as the saturation value can meet the requirements. This is reasonable as long as the signals are chosen randomly.

\subsection{Neural network selection}

Four specific ANN structures were investigated. To be useful in practice the ANN must meet two important requirements, namely the output and ¦Ã signal waveform must be identical in shape and as close as possible in magnitude. Therefore, after naked eye judges that the ANN output has a close resemblance in shape, the value of the difference between the maximum value of the output and original waveform expressed as a proportion of the original waveform is computed as shown in Eq. \ref{Equation_2} and used as an accuracy criterion.

\begin{equation}\label{Equation_2}
Err=\frac{(Max_{ori}-Max_{out})}{Max_{ori}}
\end{equation}

\subsubsection{BP neural network}

BPNNs are the most wildly used neural networks. They have a hierarchical feed forward structure in which the outputs of each layer are sent directly to every neuron of the next layer. At least three layers are present but sometimes several more. In each case an input layer distributes signals to a middle, or hidden layer, the nonlinear relationships are determined, and an output layer produces calculated data \cite{Ref_6}.

\subsubsection{Elman neural network}

The Elman NN is based on the BPNN but also includes a feedback system. The version used in the present study is described as a partial recurrent network, and lies somewhere between a classic feed-forward perception network and a pure recurrent network. The feed-forward loop consists of an input layer, a hidden layer, and an output layer in which the weights connecting two neighboring layers can be varied. In contrast to the traditional feed-forward loop, the back-forward loop includes a context layer that is sensitive to the history of input data and with fixed connections between the context layer and hidden layer \cite{Ref_7}.

\subsubsection{RBF neural network}

The RBFNN is designed to deal with intractable irregularity in the system, generally converges quickly and has been successfully applied to approximate non-linear functions, time series analysis, data classification, pattern recognition, information processing, image processing, system modeling, control and fault diagnosis. RBFNN have the strong advantage of being simple to design, are generalizable, robust and tolerant of input noise \cite{Ref_8}.
\subsubsection{GRBF neural network}

Compared to the RBFNN, the GRBFNN has advantages of requiring less computation, high model accuracy, avoids local minima during training, fast convergence and robustness \cite{Ref_9, Ref_10} and is predicted to be the ANN most suited to solving the problem of reconstructing the saturated $\gamma$ signals.

\section{Results}

The performance of the various ANNs to recover the saturated $\gamma$ signal waveforms are presented below.

\subsection{Result of BP neural network}

The output from the BP ANN does not have the shape of a $\gamma$ signal waveform, the connection between the sampling points is not smooth, and there are various discontinuities and random fluctuations. The weighting of the BPNN can be adjusted during training but it is difficult to find the most appropriate value.

\begin{figure}
 \centering
 \includegraphics[scale=0.5]{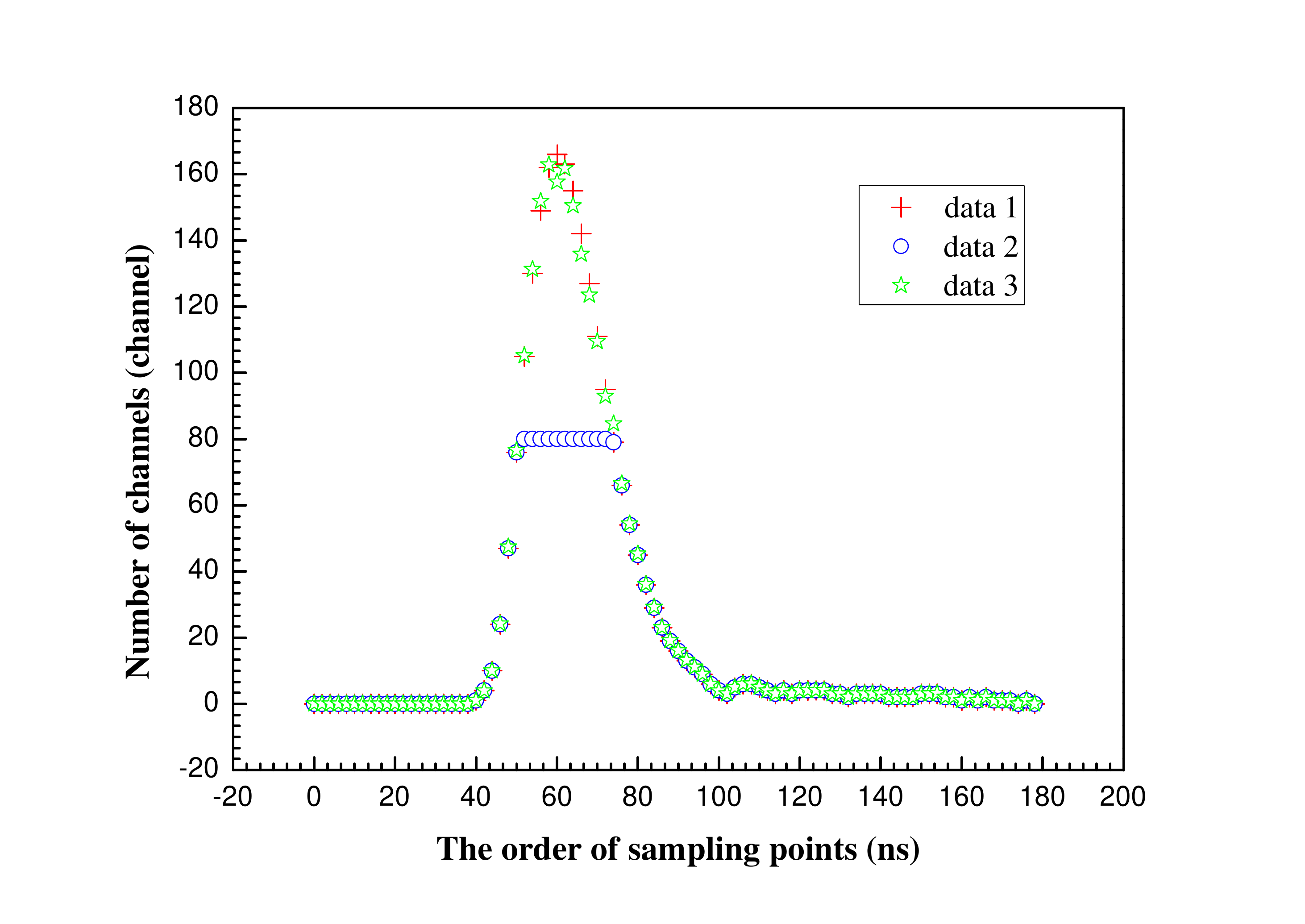}\\
 \caption{Performance of BP neural network.}\label{BP_p}
\end{figure}

The result of applying the BPNN to recover the saturated $\gamma$ signal waveforms in a set of data arbitrarily selected from the testing set is shown in Fig. \ref{BP_p} The output is not consistent with the shape of the original $\gamma$ signal especially near the maximum value.

\subsection{Result of Elman neural network}

The Elman NN showed better performance than the BPNN, but is still not consistent with the shape of the original $\gamma$ signal. Some improvements in performance can be obtained by modifying the network parameter but the difference in peak position remains. Further improvements may be possible but there is no theory to guide adjustments and the convergence speed of the Elman NN is unfortunately slow.

\begin{figure}
 \centering
 \includegraphics[scale=0.5]{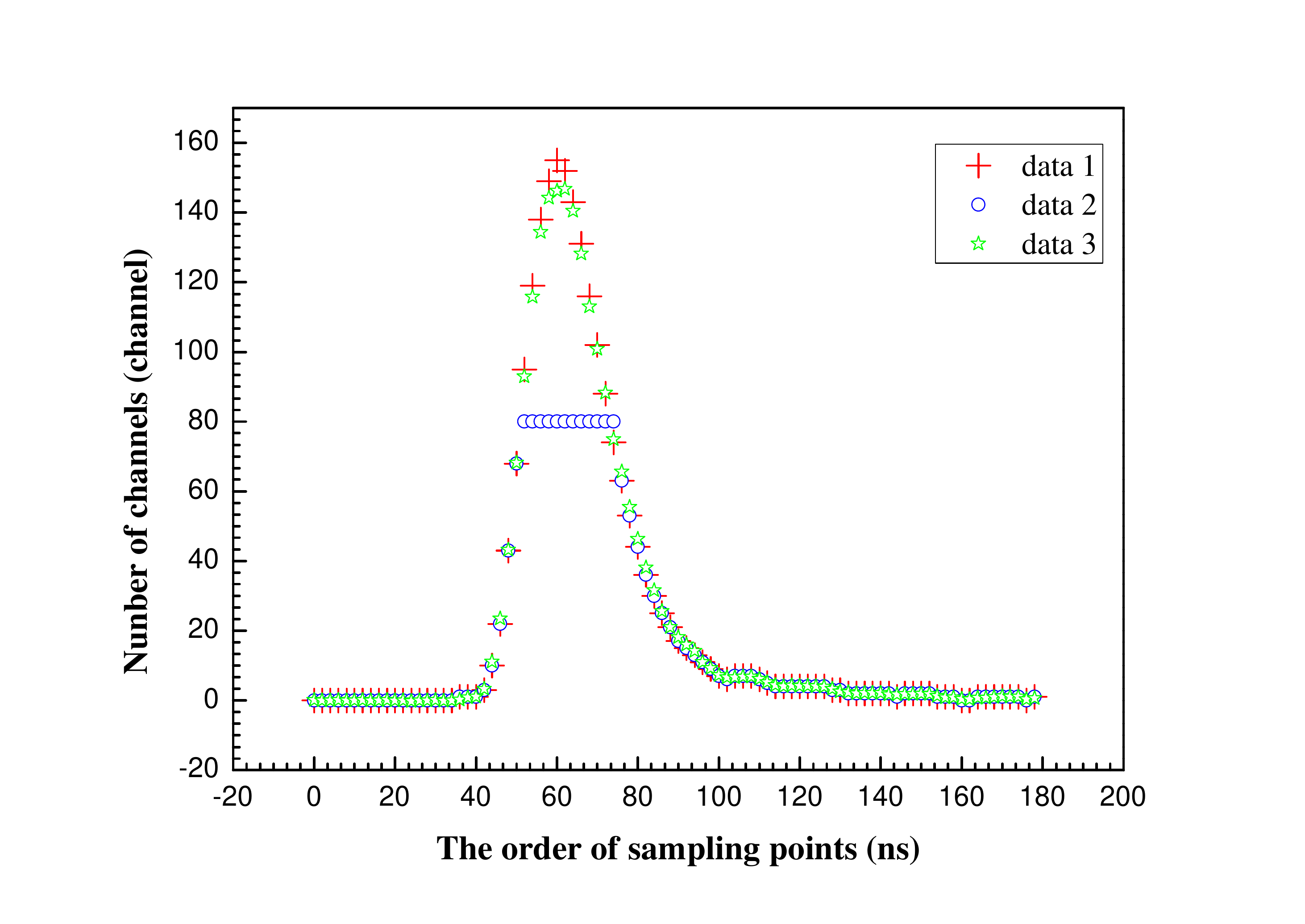}\\
 \caption{Performance of Elman neural network.}\label{Elman_p}
\end{figure}

The output of the Elman NN applied to recover the saturated gamma signal waveforms is shown in Fig. \ref{Elman_p}.There are irregularities especially near the peak values.

\subsection{Result of RBF neural network}

For the RBFNN, the output again does not have the shape of the original $\gamma$ signal waveform and could not be optimized by modifying the network parameters.

\begin{figure}
 \centering
 \includegraphics[scale=0.5]{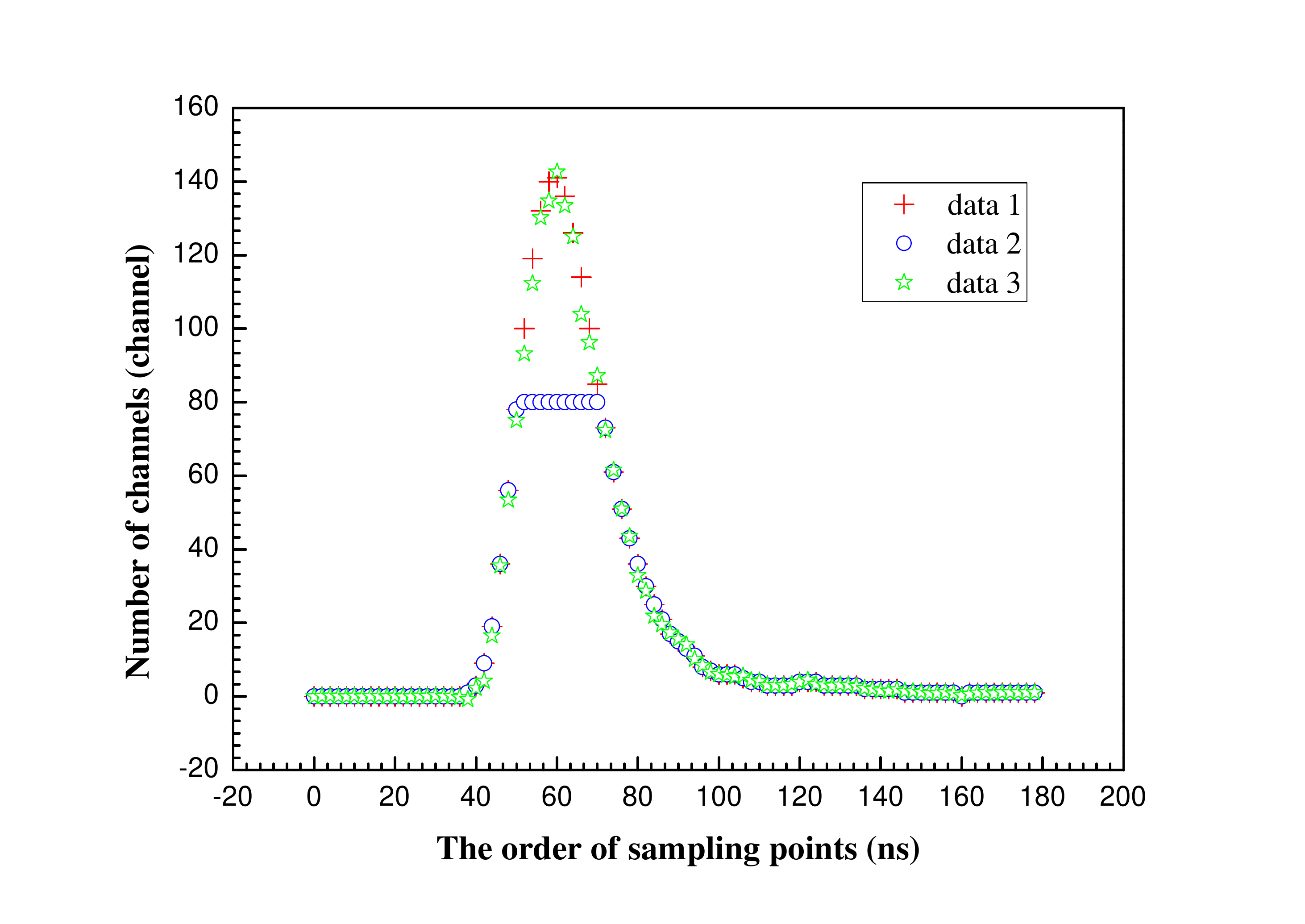}\\
 \caption{Performance of RBF neural network.}\label{RBF_p}
\end{figure}

The output of the RBFANN applied to recover the saturated $\gamma$ signal waveforms in the test dataset is shown in Fig. \ref{RBF_p} and can clearly be seen to be unsuitable.

\subsection{Result of GRBF neural network}

\begin{figure}
 \centering
 \includegraphics[scale=0.5]{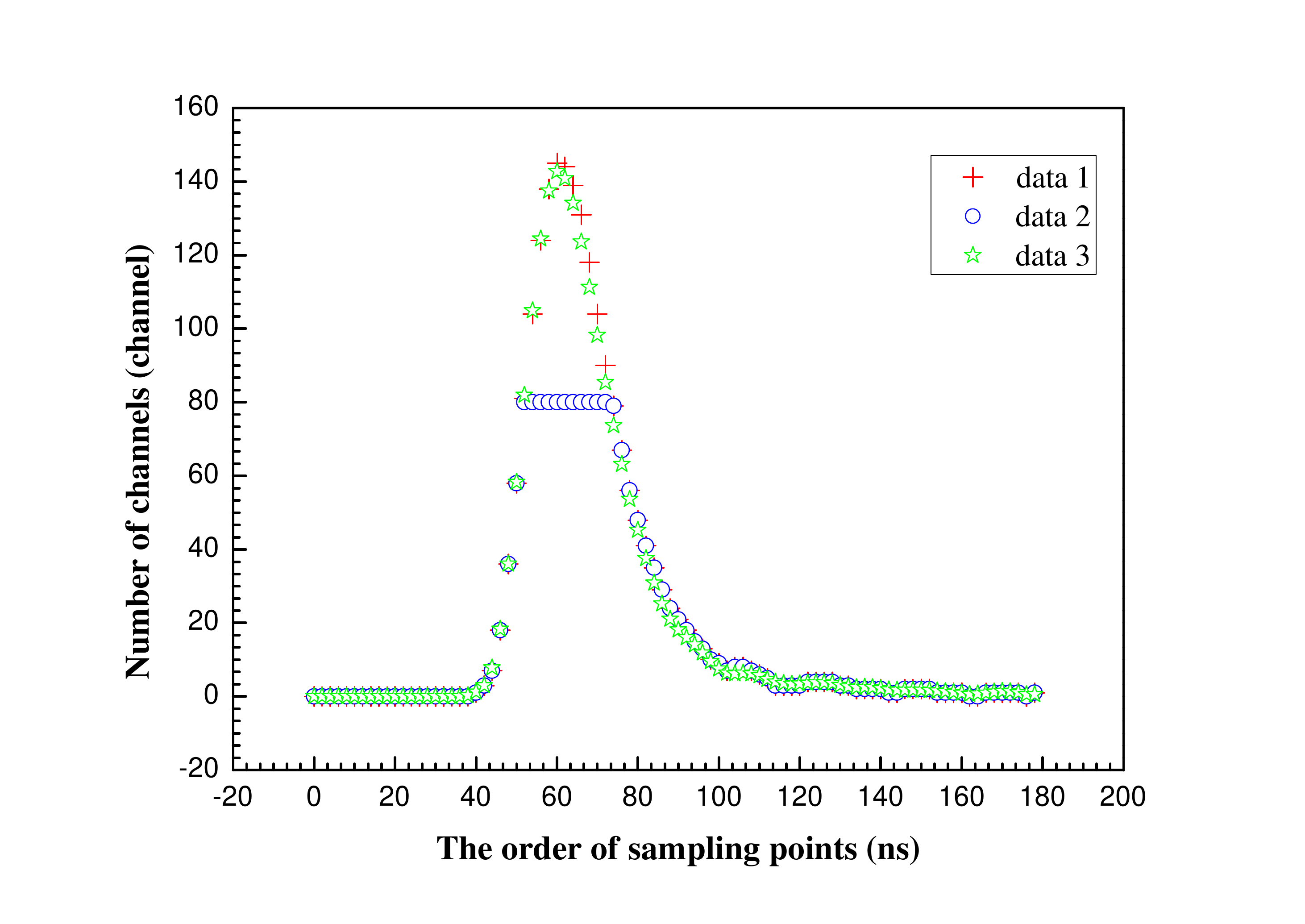}\\
 \caption{Performance of GRBF neural network.}\label{GRBF_p}
\end{figure}

\begin{figure}
 \centering
 \includegraphics[scale=0.5]{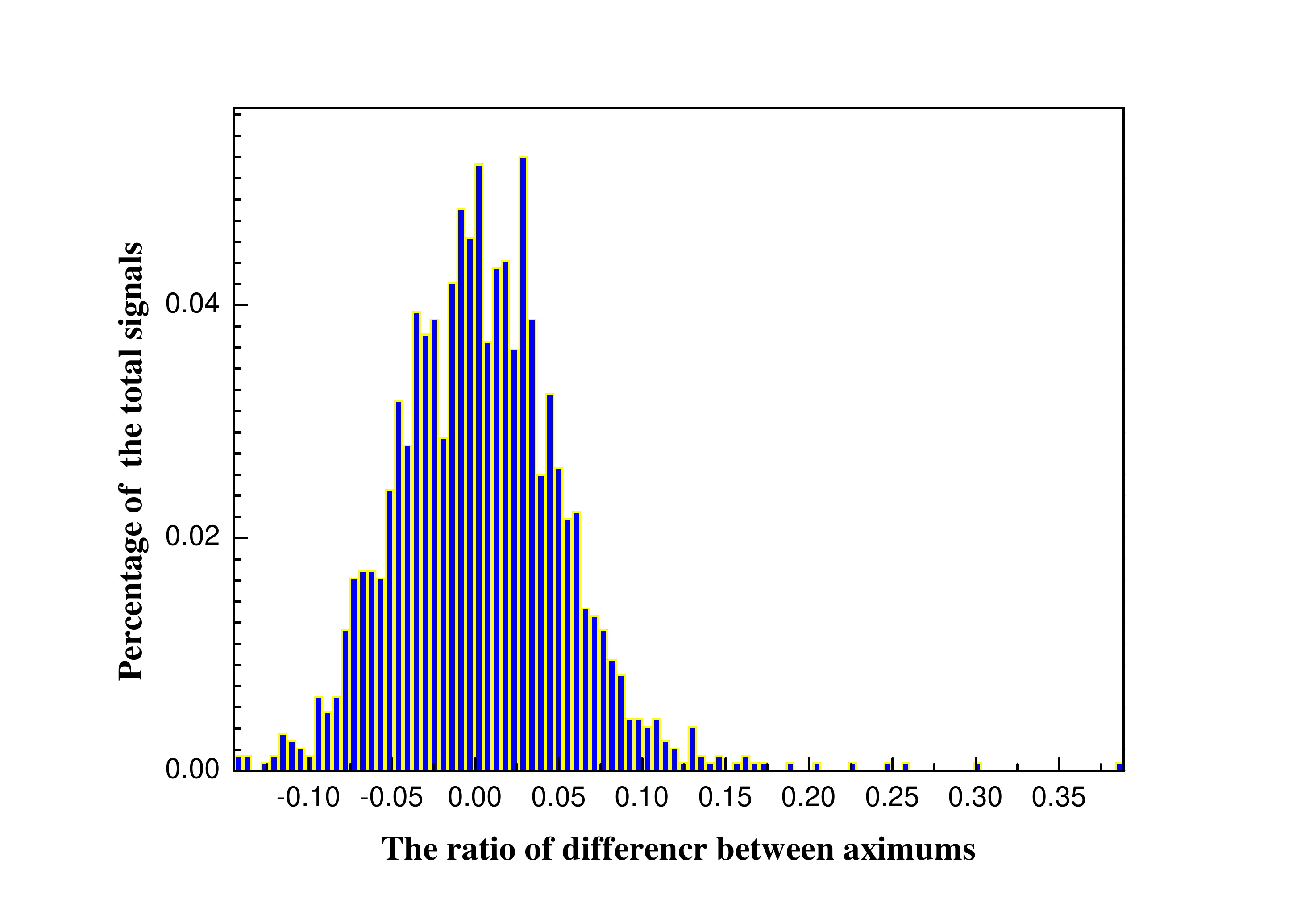}\\
 \caption{The distribution of error of the GRBF neural network.}\label{distribution_p}
\end{figure}

The GRBFNN out performs the BP, Elman and RBF NNs and the fitting curve is essentially identical in shape to the original $\gamma$ signal waveform and the output of the GBRFNN applied to recover the saturated $\gamma$ signal waveforms in the test dataset is shown in Fig. \ref{GRBF_p} and the error distribution is shown in Fig. \ref{distribution_p}. The only exception is that the maximum value is too high. We initially determined that the GRBF neural network is applicable. Then we began to modify the network parameters in order to get the desired results.

The GRBFNN is successful and suitable for restoring the saturated $\gamma$ signal waveform.

\section{Discussion and Conclusion}

In summary, the BPNN and RBFNN could not be used to successfully restore the saturated $\gamma$ signal waveforms and the Elman NN required substantial debugging and was time consuming. However, the GRBFNN was found to have excellent performance. Of course, the saturated $\gamma$ signals recovered in our tests were not yet the true saturated $\gamma$ signals detected in our experiments, but simulated saturation signals developed for the purpose of testing the approach and identifying the best performed ANN. The next stage is to select a set of normal waveforms, use them as the outputs of the training set, their truncated analog saturated waveforms as inputs for the training set, and the real saturated waveforms as inputs of the testing set. Then the outputs of the testing set are the waveforms recovered from the true saturated $\gamma$ signal. Several ANNs have been tested for recovering the saturated $\gamma$ signal waveforms recorded by liquid scintillation detectors. Since the waveforms are not linear, simple scaling cannot be used to recover the saturated signals and the Marrone model does not provide a good fit to the $\gamma$ signal waveforms. Thus four kinds of ANNs were investigated and the GRBFNN had excellent performance and allowed high energy signal waveforms to be recovered for existing hardware providing a more accurate recording for further analysis.


\section*{References}

\begin{thebibliography}{99}

\bibitem{Ref_1}
Cai-Xun Zhang, Shin-Ted Lin, Jian-Ling Zhao, Xun-Zhen Yu, Li Wang, Jing-Jun Zhu, Hao-Yang Xing,Discrimination of neutrons and ¦Ã-rays in liquid scintillator based on Elman neural network, \emph{Chin. Phys. C} 40 (2016) doi: 10.1088/1674-1137/40/8/086204

\bibitem{Ref_2}
Seyed, Abolfazl, Hosseini, Neutron spectrum unfolding using artificial neural network and modified least square method, \emph{Radiation Physics and Chemistry} 126 (2016) 75-84.

\bibitem{Ref_3}
A. Yadollahi, E. Nazemi, A. Zolfaghari, A.M. Ajorloo, Optimization of thermal neutron shield concrete mixture using artificial neural network, \emph{Nuclear Engineering and Design} 305 (2016) 146-155.

\bibitem{Ref_4}
Marrone. S, Cano-Ott. D, Colonna. N, Domingo. C, Gramegna. F, Gonzalez. E.M, Gunsing. F, Heil. M, K?ppeler. F, Mastinu. P.F, Milazzo. P.M, Papaevangelou. T, Pavlopoulos. P, Plag. R, Reifarth. R, Tagliente. G, Tain. J.L, Wisshak. K,  Pulse shape analysis of liquid scintillators for neutron studies, \emph{Nuclear Inst. and Methods in Physics Research.A} 490(1)(2002) 299-307.

\bibitem{Ref_5}
Maria GraziaBonelli, Mauro Ferrini, Andrea Manni, Artificial neural networks to evaluate organic and inorganiccontamination in agricultural soils, \emph{Chemosphere} 186 (2017) 124-131.
\bibitem{Ref_6}
Guolin Jing, Wenting Du, Yingying Guo, Studies on prediction of separation percent in electrodialysis process via BP neural networks and improved BP algorithms, \emph{Desalination} 291 (2012) 79-93.
\bibitem{Ref_7}
Sastry PS, Santharam G and Unnikrishnan KP, Memory neuron networks for identification and control of dynamic systems,\emph{IEEE Trans Neural Networks} 5 (1994) 306-319.
\bibitem{Ref_8}
Yu H, Xie T, Paszczynski S, Wilamowski BM, Advantages of radial basis function networks for dynamic system design, \emph{IEEE Transaction} 58 (2011) 5438-5450.
\bibitem{Ref_9}
W U Y M, W ANG Y Q, LI L, Application study on BP network and generalized RBF network in estimating distribution model of mechanical products[J], \emph{China Mechanical Engineering} 20 (2006) 1012-1019.
\bibitem{Ref_10}
DANG K F, YANG L B, LIN T Q, A new type of general RBF neural network and its training method, \emph{Computing Technology and Automation} 26(1) (2007) 9-13.

\end{thebibliography}
\end{document}